\documentclass[twocolumn,showpacs,preprintnumbers,amsmath,amssymb]{revtex4}
\usepackage{graphicx}% Include figure files
\usepackage{dcolumn}% Align table columns on decimal point
\usepackage{bm}% bold math
\usepackage{hyperref}% makes the references hyperlinked
\newcommand{\hodge}[1]{\,*#1}

\begin{document}
\preprint{}
\title{Compactification in first order gravity}
\author{Rodrigo Aros} \affiliation{Departamento de Ciencias F\'isicas,
\\ Universidad Andr\'es Bello, Avenida Rep\'ublica 252, Santiago,Chile}
\author{Mauricio Romo} \affiliation{Departamento de F\'isica,
Facultad de Ciencias F\'isicas y Matem\'aticas, Universidad de
Chile, Avenida Blanco Encalada 2003, Santiago, Chile}
\author{Nelson Zamorano} \affiliation{Departamento de F\'isica,
Facultad de Ciencias F\'isicas y Matem\'aticas, Universidad de
Chile, Avenida Blanco Encalada 2003, Santiago, Chile}
\date{\today}

\begin{abstract}
The Kaluza-Klein compactification process is applied in five
dimensions to Chern Simons gravity, for the anti-de Sitter and
Poincar\'e groups, using the first order formalism. In this context
some solutions are found and analyzed. Also, the conserved charges
associated to the solutions are computed.

\end{abstract}

\maketitle

\section{Introduction}

The Kaluza-Klein construction \cite{Kaluza:1921tu,Klein:1926tv}
showed that four dimensional interactions could be understood as
manifestations of an underlying higher dimensional gravity. In five
dimensions the gravitational theory is defined over a manifold
$\mathcal{M}_{5}$, with the topology of
$\mathcal{M}_{5}=\mathcal{M}_{4}\times S^{1}$ where
$\mathcal{M}_{4}$ is a four dimensional manifold. This construction
actually corresponds to study the \textit{gravity} of a fiber bundle
\cite{yvonne} where $S^{1}$ and $\mathcal{M}_{4}$ are the fiber and
the base space respectively.

The geometry of $\mathcal{M}_{5}$ motivates the introduction of a
coordinate system $(x^{\mu},\varphi)$, with $x^{\mu}$ the
coordinates on $\mathcal{M}_{4}$ and $\varphi\in[0, 2\pi[$ in order
to describe the $S^{1}$. In principle, one should introduce a
Fourier expansion in $\varphi$ for every field on $\mathcal{M}_{5}$,
however only the lowest order will be considered here. With these
coordinates the line element at \textit{lower order in the fifth
dimension} reads
\begin{eqnarray}
ds^{2}_{5} &=& \left(g_{\mu\nu}(x) + \Phi(x)^{2}A_{\mu}(x) A_{\nu}(x)\right) dx^{\mu} dx^{\nu} \nonumber\\
&+&2\Phi(x)^{2} A_{\mu}(x) dx^{\mu}d\varphi +
\Phi(x)^{2}d\varphi^{2}\label{LowerMetric},
\end{eqnarray}
$A_{\mu}(x)$ is identified with an electromagnetic field potential
and $\Phi(x)$ with a scalar field on $\mathcal{M}_{4}$.

On the other hand, the introduction of fermions into gravity drags
the need of extending the metric gravity making manifest the
presence of a local Lorentz group in the tangent space
\cite{VanNieuwenhuizen:1981ae}. To achieve this one needs to
introduce an orthonormal basis $\tilde{e}^{A}$, usually called
vielbein, and a connection for the local Lorentz group,
$\tilde{\omega}^{AB}$, called spin connection. If the spin
connection is considered an independent field then a reformulation
of gravity, called first order gravity, arises. This formulation has
proven to be worthy on its own, beyond the presence of fermions.

In this work some aspects of compactification of first order gravity
are addressed. In particular some solutions are shown as well as
their analysis. Compactification of a first order theory of gravity
differs from metric formalism and allows to visualize some aspects
which are usually ignored, for instance, the presence of torsion in
$\mathcal{M}_{4}$. Given that Einstein theory has been well studied
within Kaluza-Klein construction this work concentrate mostly on
Chern Simons (CS) gravities.

\section{Gravity and compactification}
To begin with the discussion, the five dimensional case, shown in
Eq.(\ref{LowerMetric}), will be reanalyzed in the context of first
order gravity. To obtain the metric (\ref{LowerMetric}) one can
choose general f\"unfbein $\tilde{e}^{A}$, with $A=0\ldots 3,5$,
\begin{equation}\label{vielbein} \tilde{e}^{a} = e^{a}(x)  \textrm{ and } \tilde{e}^{5} =
\Phi(x)(A(x) + e^{5})
\end{equation}
with latin index $a=0\ldots 3$, $\Phi(x)$ is a scalar field, $A(x)$
is a 1-form on $\mathcal{M}_{4}$ and $e^{5}=d\varphi$.

To introduce a connection compatible with the vielbein above one
have to consider that $\xi=\partial_{\varphi}$ is a Killing vector
for Eq.(\ref{LowerMetric}). A Killing vector generates a Lorentz
transformation with parameters $ \Delta^{AB}=
I_{\xi}\tilde{\omega}^{AB} - \tilde{E}^{A M} \tilde{E}^{B N}
(\nabla_{M} \xi_{N})$, however in this case, see
Eq.(\ref{vielbein}), $\Delta^{AB}(\xi)=0$. Using this result, the
most general connection compatible with Eq. (\ref{LowerMetric})  is
given by
\begin{equation}\label{SpinConnection}
     \tilde{\omega}^{ab} = \omega^{ab}(x) + \psi^{ab}(x) e^{5}\textrm{ and } \omega^{a5} = v^{a}(x) +
     p^{a}(x)
     e^{5},
\end{equation}
where $\psi^{ab}(x)$ and $p^{a}(x)$ are scalars and $\omega^{ab}(x)$
and $v^{a}(x)$ are a one-form respectively on $\mathcal{M}_{4}$.

The vielbein (\ref{vielbein}) and the connection
(\ref{SpinConnection}) determine the curvature
$\tilde{R}^{AB}=d\tilde{\omega}^{AB}
+\tilde{\omega}^{A}_{\hspace{1ex}C}\tilde{\omega}^{CB}$, obtaining
\begin{eqnarray}
   \tilde{R}^{ab} &=&  (R^{ab} - v^{a} v^{b}) + (D(\psi^{ab}) + p^{a}
v^{b}- p^{b} v^{a}) e^{5},\nonumber\\
   \tilde{R}^{a5} &=&  D(v^{a}) + (Dp^{a} - \psi^{a}_{\hspace{1ex} b}
e^{b}) e^{5},\label{Curvature}
\end{eqnarray}
and the torsion two-form $\tilde{T}^{A}=d\tilde{e}^{A}+
\tilde{\omega}^{A}_{\hspace{1ex}B}\tilde{e}^{B}$
\begin{eqnarray}
   \tilde{T}^{a} &=& (T^{a} - \Phi v^{a} A) - (\psi^{a}_{\hspace{1ex} b} e^{b}-A\Phi p^{a} + v^{a}\Phi) e^{5},\nonumber\\
   \tilde{T}^{5} &=& (d(\Phi A) - v_{c} e^{c})+(d\Phi + p_{c} e^{c}) e^{5}.\label{Torsion5} \end{eqnarray}

\section{Definitions of charges}

Because the Kaluza-Klein construction is a very particular geometry,
one can address part of the analysis of charges on  a general ground
without considering a particular theory of gravity.

The form of the Lagrangian,
\begin{equation}\label{LagragianForm}
L_{5} = \left(L_{4}(x) + dB_{3}(x)\right)\wedge d\varphi
\end{equation}
guaranties that the Noether charges obtained in five dimensions from
$L_{5}$ are connected with the Noether charges in four dimensions
from $L_{4} + dB_{3}$ by
\begin{equation}\label{ChargesInCharges}
Q = \int_{\partial \Sigma_{3}\times S^1} \hodge{\mathbf{J}}_{5}
\equiv 2\pi\int_{\partial \Sigma_{3}} \hodge{\mathbf{J}}_{4},
\end{equation}
where $\Sigma_{3}$ represents a family of space-like surfaces that
foliates $\mathcal{M}_{4}$. In this way the effective action in four
dimensions contains all the physics of five dimensions. After this
remark it becomes straight to obtain the mass or the angular momenta
of any solution of this theory as the Noether charges associated
with Killing vectors on $\mathcal{M}_{4}$.

Recalling that by construction $\xi=\partial_{\varphi}$ is a Killing
vector, the analysis above can be extended to obtain the electric
charge. In any electromagnetic theory, the electric charge can be
obtained as the Noether charge associated with the gauge
transformations, $Q(\lambda)$, whose gauge parameters, say
$\lambda(x)$, can be smeared out at infinity, in this case at
$\partial \Sigma_{3}$. Thus the electric charge is given by
\begin{equation}\label{Electricchargedef}
     \tilde{q} = \left(\frac{1}{\lambda_{0}}
     Q(\lambda)\right)_{\partial\Sigma_{3}}
\end{equation}
with $\left.\lambda(x) \right|_{\partial \Sigma_{3}}=\lambda_{0}$.
On the other hand, The Kaluza-Klein construction
Eqs.(\ref{vielbein},\ref{SpinConnection}) is invariant under the
transformation,
\begin{eqnarray}
   \varphi &\rightarrow& \varphi + \lambda(x)\label{gaugetransformation} \\
   A(x) &\rightarrow& A(x) + d\lambda(x) \nonumber,
\end{eqnarray}
where one recognizes a gauge transformation of $A$. By noticing that
the subset of gauge transformations useful for
Eq.(\ref{Electricchargedef}) coincides with the transformation
generated by $\xi=\partial_{\varphi}$, one finally obtains
\begin{equation}\label{PrimordialElectricCharge}
 q_{4} = \int_{\partial \Sigma_{3}\times S^1} \hodge{\mathbf{J}}_{5}(\xi).
\end{equation}

\section{Einstein gravity}

The five dimensional (first order) Einstein Hilbert (EH) action
reads
\begin{equation}\label{EH}
I_{EH} = \kappa_{G} \int_{\mathcal{M}_{5}} \tilde{R}^{AB}
\tilde{e}^{C} \tilde{e}^{D} \tilde{e}^{F}\varepsilon_{ABCDF}.
\end{equation}
It yields the equations of motion
\begin{eqnarray}\label{EinsteinEquations}
\mathcal{E}_{F} &=& \tilde{R}^{AB} \tilde{e}^{C} \tilde{e}^{D}
\tilde{e}^{F}\varepsilon_{ABCDF}=0,\nonumber\\ \tilde{T}^{A} &=& 0
\end{eqnarray}
It must be stressed that in first order gravity the vanishing of
torsion is a consequence of the equations of motion.

The vanishing of torsion, $\tilde{T}^{A}=0$, by Eq. (\ref{Torsion5})
determines that
\begin{equation}\label{Elect1}
v^{a} = -E^{a\mu}\left(\frac{1}{2}\Phi F_{\mu\nu}dx^{\nu}  +
\partial_{\mu}\Phi A\right),
\end{equation}
\begin{equation}\label{Elect2}
p_{a}=-E_{a}^{\hspace{1ex} \mu} \partial_{\mu} \Phi \textrm{ and
}\psi_{ab} = -\frac{1}{2}E_{a}^{\hspace{1ex} \mu}
E_{b}^{\hspace{1ex}\nu} F_{\mu\nu},
\end{equation}
with $F_{\mu\nu}=\partial_{\mu}A_{\nu}-\partial_{\nu}A_{\mu}$. This
last equation identifies $A$ with a field potential.

Remarkably the four dimensional torsion (see Eq.(\ref{Torsion5}))
not only does not vanish, but it actually reads
\begin{equation}\label{Contorsion}
T^{a} = \Phi A v^{a} \Leftrightarrow K^{ab} = \Phi F^{ab} A,
\end{equation}
where $K^{ab}$ is the contorsion one form.

In metric formalism one can skip the presence of torsion because it
can be completely understood in terms of an electromagnetic field,
see Eq.(\ref{Contorsion}). This feature is not surprising, in fact
it is well established that under some particular conditions, as
those given here, a torsion tensor can be rewritten as an effective
electric field, although this have been proven to be false in
general \cite{DeChingChern}.

\subsection*{$\Phi$ can not be constant}
After the replacement of conditions (\ref{Elect1}) and
(\ref{Elect2}) -obtained from $\tilde{T}^{A}=0$- into the action
(\ref{EH}) it becomes
\begin{equation}\label{EffectiveEHElec}
I_{EH}=\kappa_{G}\int_{\mathcal{M}_{4}\times S^{1}}\Phi \left( R +
\frac{1}{4}\Phi^{2}F_{\mu\nu} F^{\mu\nu}\right) \sqrt{g}d\varphi
d^4x,
\end{equation}
where $R$ is the standard four dimensional Ricci scalar. Since this
five dimensional action (\ref{EffectiveEHElec}) is independent of
$\varphi$ one can integrate it out obtaining the effective action
\begin{equation}\label{EffectiveEHElecII}
I_{eff} =2\pi\kappa_{G}\int_{\mathcal{M}_{4}} \Phi \left( R +
\frac{1}{4}\Phi^{2}F_{\mu\nu}F^{\mu\nu}\right)\sqrt{g}\,d^4x.
\end{equation}
It must be stressed that the equations of motion obtained from this
action reproduce the Einstein equations (\ref{EinsteinEquations})
after using the ansatz (\ref{vielbein}) and (\ref{SpinConnection}).

Observing the final expression (\ref{EffectiveEHElecII}) one could
consider to take $\Phi$ constant, and so to obtain the standard
Einstein Maxwell theory. However this breaks the equivalence between
five and four dimensions, since a constant $\Phi$, through
$G^{\varphi}_{\hspace{1ex} \varphi}=0$, implies $F_{\mu\nu}=0$,
yielding a trivial result.

\section{Beyond Einstein}

In higher dimensions the premise of second order equation of motion
for the metric does not restrict the action to EH. One the simplest
extension gives rise to Lovelock gravities
\cite{Lovelock:1971yv,Zumino:1985dp,Zanelli:2002qm}. Furthermore in
odd dimensions there are subfamilies of Lovelock gravities which
coincide with CS gravities and so non vanishing torsion solutions
\cite{Chamseddine:1989nu} exists. Unfortunately only a few non
vanishing torsion solutions are known
\cite{Banados:2001hm,Aros:2006qc} at present time. It is worth to
stress that CS gravities are genuine gauge theories for (A)dS and
Poincar\'e groups respectively.

The Poincar\'e CS action in five dimensions \cite{Banados:1996hi} is
the so called \textit{Gauss Bonnett} term and reads
\begin{equation}\label{CS}
     I_{p}=\kappa_{G} \int_{\mathcal{M}_{5}} \tilde{R}^{AB} \tilde{R}^{C D} \tilde{e}^{F}\varepsilon_{ABCDF} +
     dB_{4},
\end{equation}
where  $B_{4}(x)$ is boundary term to be fixed later. Its equations
of motion are $ \tilde{R}^{AB} \tilde{R}^{C D} \varepsilon_{ABCDF} =
0 $ and $ \tilde{R}^{AB} \tilde{T}^{C} \varepsilon_{ABCDF}= 0$.

\subsection*{Effective theory}
Using the Kaluza-Klein ansatz in the five dimensional action
(\ref{CS}) together with the vanishing of torsion one can compute an
effective four dimensional action starting from Eq.(\ref{CS}). The
effective action reads
\begin{equation}\label{Eff4d}
I_{eff}=2\pi\kappa_{G}\int_{\mathcal{M}_{4}}\Phi\left(\Omega^{ab}\Omega^{cd}+4\rho^{a}\tau^{bc}e^{d}+4\sigma^{a}\Omega^{bc}e^{d}\right)\varepsilon_{abcd}
\end{equation}
where
\begin{eqnarray}
   \Omega^{ab} &=&  \hat{R}^{ab} - \frac{1}{2}\Phi^{2}F^{ab}F- \frac{1}{4}\Phi^{2}F^{a}_{\hspace{1ex} c}F^{b}_{\hspace{1ex} d}e^{c}e^{d} \nonumber\\
   \tau^{ab} &=&  -\frac{1}{2}\hat{D}(\Phi^{2}F^{ab}) + \frac{1}{2}d\Phi F^{ab} + \partial^{[a}\Phi
   F^{b]}_{\hspace{1ex} c}e^{c} \nonumber\\
   \rho^{a} &=&  \frac{1}{2}\hat{D}(\Phi F^{a}_{\hspace{1ex} b})e^{b}
+ \partial^{a}\Phi F \nonumber\\
   \sigma^{a} &=&  \frac{\hat{D}(\partial^{a}\Phi)}{\Phi} + \frac{1}{4}\Phi^{2}F^{a}_{\hspace{1ex} c}F^{c}_{\hspace{1ex}
   d}e^{d}\label{Effectivefields}
\end{eqnarray}
The $\hat{\,}$'s on the derivative indicates that they are
torsionless derivatives on four dimensions, \textit{i.e.}, the
contorsion $K^{ab}$ has been explicitly separated in the equations
above.

The equations of motion, written in terms of the effective fields
displayed above, Eqs.(\ref{Effectivefields}), are cumbersome, thus
we chose not to write them down. The action (\ref{Eff4d}) reproduces
the five dimensional CS equations of motion. It is straightforward
to prove that a constant $\Phi$, just as before, implies the
vanishing of $F_{\mu\nu}$.

\section{Solution}

In this section some solutions of CS gravities in five dimensions
will be discussed. A solution of the Poincar\'e CS gravity with
spherical symmetry in four dimensions is given by
\begin{eqnarray}
   e^{0}=N(r)dt, & & e^{1} =  \frac{1}{g(r)}dr,  \nonumber\\
   e^{2} =  r d\theta, & &  e^{3} = r\sin(\theta)d\phi, \label{vielbeinsolution}\\
   \Phi = \Phi(r), & & A= a(r)dt,\nonumber \end{eqnarray} where \begin{eqnarray}
   \Phi(r) &=&
c_{1}r\pm\sqrt{c_{1}^{2}r^{2}+c_{2}r+c_{3}}+\frac{c_{2}}{2c_{1}},
\nonumber\\
   N(r) &=& g(r) =
\sqrt{1-\frac{8q^{2}c_{1}^{3}}{(c_{2}^{2}-4c_{1}^{2}c_{3})}\left(\frac{d\Phi}{dr}\right)^{-1}}
,\nonumber\\
   a(r) &=&  \frac{q}{\Phi^{2}}.\label{explicitfields}
\end{eqnarray}

In this solution one can recognize four arbitrary integration
constants, which occurs because CS gravity has non linear equations
of motion.

\subsection{Analysis}
The analysis of the four dimensional metric is best carried out
using the variable $R=r+\frac{c_{2}}{2c_{1}^2}$. So the metric is
written \begin{equation}
ds^{2}_{4}=-N(R)^{2}dt^{2}+\frac{1}{N(R)^{2}}dR^{2}+\left(R-\frac{c_{2}}{2c_{1}^2}\right)^{2}d\Omega^{2}
\end{equation}
\begin{eqnarray}
   \Phi(R) &=& c_{1}\left(R+sgn(c_{1})\sqrt{R^{2}+\kappa}\right), \nonumber\\
   N(R) &=&
\sqrt{1+\frac{2q^{2}}{c_{1}\kappa}\left(\frac{d\Phi}{dR}\right)^{-1}}
,\nonumber\\
   a(R) &=&  \frac{q}{\Phi(R)^{2}}
\end{eqnarray}
with
\begin{equation}
     \kappa = \frac{4c_{1}^{2}c_{3}-c_{2}^{2}}{4c_{1}^{4}}
     \nonumber
\end{equation}

The values $\kappa<0$ and $c_{2}>0$ lead to naked singularities or a
metric with the wrong signature everywhere, therefore they are
dismissed from the physical spectrum. The case with $c_{1}<0$,
$\kappa<0$ and $c_{2}=0$ deserves some attention and will be
analyzed in a subsequent section.

The case $\kappa>0$, $c_{1}>0$ and $c_{2}=0$ leads to a solution
which is regular everywhere and is asymptotically flat,
\textit{i.e.},
\begin{equation} \lim_{r\rightarrow \infty} R^{\mu\nu}_{\alpha\beta}
= 0,
\end{equation} in four and five dimensions. This solution
may be regarded as a soliton. Because of its regularity this case
will analyzed in detail.

\subsection{The definitions of charges}

After realizing that the action (\ref{CS}) is a CS action for the
Poincar\'e group, one can skip the long process of reobtaining the
Noether charges. The Noether currents associated with the Killing
vectors of a CS theory have been discussed in
Ref.\cite{Mora:2006ka}. In five dimensions it is given by
\begin{equation}\label{Jdiff}
\hodge{{\bf J}_{5}}(\eta) = 6\, d\left( \int_{0}^{1}dt \langle
(A_{1}-A_{0}) F_t I _{\eta}A _t\rangle \right),
\end{equation}
where $\eta$ is a Killing vector. $F_{t}=dA_{t} + A_{t}\wedge A_{t}$
with $A_{t} = t A_{1} + (1-t) A_{0}$. $\langle \rangle$ is the trace
in the group. Here $A_{0}$ and $A_{1}$ are connections in the same
fiber having the generic form
\begin{equation}\label{A}
A = \frac{1}{2} \tilde{\omega}^{AB} J_{AB} + \tilde{e}^{A} P_{A},
\end{equation} $P_{A}$ and $J_{AB}$ are the generators of the Poincar\'e  group. The charges are computed using the background $A_{0}$. It is worth to stress that background
independent methods exists to calculate Noether charges for the
CS-AdS gravity \cite{Mora:2006ka,Mora:2004kb} but they could not be
trivially adapted to the Poincar\'e case. The definition of $A_{0}$
as a flat connection, determines $B_{4}$ in Eq.(\ref{CS}). In this
way the Noether charges are given by
\begin{equation}\label{NoetherCharges}
Q(\eta) = \int_{\partial \Sigma \times S^{1}} \hodge{{\bf
J}}_{5}(\eta).
\end{equation}

The charges of the above solution are associated with the Killing
vectors $\zeta=\partial_{t}$ and $\partial_{\varphi}$ respectively.
On the other hand $A_{0}$ will be fixed as the flat connection
obtained from the geometry, \[ ds^{2}_{bg} = -dt^{2} + dr^2 + r^2
(d\theta^2+\sin(\theta)^{2}d\phi^{2})+ d\varphi^2, \] which is a
five dimensional Minkowski space with one of its direction wrapped
up.

\subsection{Mass and electric charge}

The mass can be obtained from the five dimensional Noether charge by
Eq.(\ref{ChargesInCharges}) associated with the Killing vector
$\zeta$ using Eq.(\ref{NoetherCharges}). To compute the Noether
charge of the five dimensional CS theory is formally simpler than
the analysis in four dimensions where the effective theory, Eq.
(\ref{Eff4d}), is not purely gravitational but it contains matter.
For the case $\kappa>0$, $c_{1}>0$ and $c_{2}=0$ the mass is given
by
\begin{eqnarray}
M = Q(\zeta) &=& 8\pi^{2}\kappa_{G}
c_{1}\sqrt{1+\frac{q^2}{c_{1}^{2}\kappa}}\left(\frac{q^2}{c_{1}^{2}\kappa}\right.\nonumber
\\
&+&
\left.4\left(\sqrt{1+\frac{q^2}{c_{1}^{2}\kappa}}\left(1+\frac{9q^2}{4c_{1}^{2}\kappa}\right)-1\right)\right).\label{Mass}
\end{eqnarray}
One can check that this mass is positive. Because the mass vanishes
for $q=0$ this solution can be cast as a pure electromagnetic
solution, where the mass $M$ represents the \textit{mass} of the
electromagnetic field. This conjecture seems to be confirmed by the
asymptotic behavior of this solution, where
\[
N(r)^{2} \approx 1 + \frac{Q^{2}}{r^{2}} + \ldots
\]
reproducing the case $m=0$ and $Q\neq 0$ in the Reissner-Nordstr\o m
solution. One may speculate that another solution, one that
asymptotically behaves as RN solution with $m \neq 0$, should exist.
Unfortunately the equations of motion obtained from the action
(\ref{Eff4d}) do not allow an obvious extension of the solution
above (\ref{vielbeinsolution}) to confirm this conjeture.

In a similar way, the electric charge can be obtained using
Eq.(\ref{PrimordialElectricCharge}), where in this case the current
to be integrated is given by Eq.(\ref{Jdiff}). After a
straightforward computation the electric charge is given by
\begin{equation}\label{ElectricCharge}
     \tilde{q}=Q(\xi)=-\frac{96\pi^{2}\kappa_{G}q^3}{c_{1}\kappa}.
\end{equation}

\section{AdS in  four }

As previously noticed, the $c_{1}<0$ case requires a deeper
analysis. In this case the function $f(R)^{2}$ diverges at
$R\rightarrow\infty$ as
\begin{equation}
f(R)^{2} \sim
1+\frac{3q^{2}}{c_{1}^2\kappa}+\frac{4q^{2}}{c_{1}^2\kappa^{2}}R^{2}
\end{equation}
giving rise to an effective cosmological constant in four
dimensions.

For $\kappa>0$ the solution is regular everywhere and there is no
horizon. The solution with $\kappa<0$ has a singularity at
$R=\sqrt{-\kappa}$ as well as an horizon at
\begin{equation}
r_{+}=\frac{c_{1}^{2}\kappa+2q^{2}}{|q|}\sqrt{\frac{-\kappa}{c_{1}^{2}\kappa+q^{2}}}
\end{equation}
for the range $-\frac{q^{2}}{c_{1}^{2}}<\kappa<0$.

For all values of $\kappa$ this solution has the same electric
charge of the $c_{1}>0$ case, see Eq.(\ref{ElectricCharge}).

Unfortunately, the mass $Q(\zeta)$, diverges and furthermore it may
not be possible to find out a background which can subtracts these
divergences. Even though such a background may exists still this
would be odd, since for a theory invariant under the Poincar\'e
group only a flat space can represent a proper background. Maybe a
background independent method to calculate the mass would give a
finite one \cite{PrivateTroncoso}.

Given these considerations, the $c_{1}<0$ case must be excluded from
the physical spectrum in the solution above.

\section{Cones}\label{Cones}

Poincar\'e CS gravity restricted to a vanishing torsion solutions is
known to have solutions with conical singularities. In a certain way
this a generalization of what is well known to happen in 2+1
dimensions where the existence of black holes is only possible with
a negative cosmological constant. This feature is shown by the
famous BTZ solution whose $\Lambda\rightarrow 0$ limit yields a
cone. For that reason the analysis of conical solutions in this
Poincar\'e CS-KK model can be of interest.

One of those solutions is given by the same ansatz
(\ref{vielbeinsolution}) with $N(r)=\alpha$, $g(r)=\beta$,
$a(r)=a_{0}$ and $\Phi(r)$ arbitrary. This solution represents a
scalar field, $\Phi$, defined over a manifold with a conical
singularity.

The arbitrariness of a field is not new for the CS gravity, and
although it may seem odd it is a natural consequence of a higher
power differential operator. In fact one must note that $\alpha$ and
$\beta$ are also arbitrary.

The arbitrariness of the scalar field can be fixed by requiring that
the solution has a physical meaning. To fulfill this requirement
$\Phi$ must be smooth near $r=0$ and the mass associated to this
solution be a finite one.

As expected the electric charge vanishes in this case since
$a=a_{0}$. The mass, $M = Q(\zeta)$, is given by
\begin{equation}\label{Conemass}
M= 4\pi^{2}\kappa_{G}\frac{d
\Phi(r)}{dr}\beta(\beta-1)\left(5+7\alpha+3\beta(1+3\alpha)\right),
\end{equation}
which is finite provided
\[
\lim_{r\rightarrow \infty}\Phi(r) \approx c_{p} r + c_{q} +
O\left(\frac{1}{r}\right), \] with $c_{p}$ and $c_{q}$ constants.

This result constraints the arbitrariness of $\Phi$.

Comparing the above result Eq.(\ref{Conemass}) with the previous
solution Eq.(\ref{explicitfields}) one finds that $\beta=1$ is
equivalent to $q=0$. That both solutions above share a sub-sector it
probably indicates that there is a more general solution that
includes both as particular cases.

\section{AdS in five}

The introduction of a cosmological constant, $\Lambda\neq 0$, into a
compactification procedure is not straightforward. First, one has to
consider that if a natural ground state for $\Lambda=0$ is a flat
space with a non vanishing cosmological constant the ground state is
expected to be a constant curvature manifold. An ansatz of the form
$AdS_{4}\times S^{1}$ don't fulfill this condition. Roughly
speaking, this implies that the fifth dimension in
Eq.(\ref{LowerMetric}) needs a \textit{warp} factor, represented by
a non constant $\Phi$, even for the ground state.

The description of a fiber bundle by a constant curvature manifold
can be difficult. One needs to isolate a cycle in $\mathcal{M}_{5}$
which can be identified with the fiber. Fortunately for the case of
negative curvature manifolds there are a plethora of known spaces
obtained as identifications of AdS which have a cycle by
construction.

The extension of Einstein gravity with a negative cosmological
constant is direct. For this reason only the AdS CS gravity will be
discussed. This gravity is given by
\begin{eqnarray} I_{p} &=& \kappa_{G} \int_{\mathcal{M}_{5}} \left(\tilde{R}^{AB} \tilde{R}^{C D}
\tilde{e}^{F} + \frac{2}{3l^{2}}\tilde{R}^{AB} \tilde{e}^{C}
\tilde{e}^{D}
\tilde{e}^{F}\nonumber\right.\\
&+&\left. \frac{1}{5l^{4}} \tilde{e}^{A}\ldots
\tilde{e}^{F}\right)\varepsilon_{ABCDF}+ dB_{4}\label{CSnegative},
\end{eqnarray} where $B_{4}$ is a boundary term to be defined later.
The corresponding equations of motion are $\bar{\tilde{R}}^{AB}
\bar{\tilde{R}}^{C D} \varepsilon_{ABCDF} = 0 $ and $
\bar{\tilde{R}}^{AB} \tilde{T}^{C} \varepsilon_{ABCDF}= 0$, where
\[ \bar{\tilde{R}}^{AB} = \tilde{R}^{AB}  + \frac{1}{l^2}\bar{e}^{A} \bar{e}^{B}.
\]
The negative cosmological constant is given by $\Lambda=-6\,l^{-2}$.
Black hole solutions for AdS CS theory can be found in
\cite{Banados:1994ur}.

\subsection*{A solution}

For simplicity one can consider to turn off the electromagnetic
field as a first approximation. In this case, to be consistent with
the fiber bundle geometry, $\mathcal{M}_{5}=\mathcal{M}_{4}\times
S^{1}$, the spherical transverse section in (\ref{vielbeinsolution})
must be replaced by a flat transverse section. After considering
this simplification a solution is given by
\begin{eqnarray}
   \Phi(r) &=& C_{1}\left(\sqrt{3\gamma + 3\frac{r^{2}}{l^{2}}}-\frac{r}{l}\right), \nonumber\\
   g(r) &=&  \sqrt{\gamma + \frac{r^{2}}{l^{2}}}\nonumber\\
   N(r) &=&  C_{1}\left(\sqrt{3\gamma + 3\frac{r^{2}}{l^{2}}}-\frac{r}{l}\right)\nonumber\\
   a(r) &=&  A_{0}\nonumber\\
   \tilde{e}^{2} &=& r d\theta\label{NegativeCurvatureSolA=0}\\
   \tilde{e}^{3} &=& rd\phi.\nonumber
\end{eqnarray}

This solution has an horizon at
\[
r_{+} = l \sqrt{-\frac{3}{2} \gamma}
\]
provided that $\gamma<0$. This solution has no meaning for $r<r_{+}$
($N(r)$ becomes complex). The inner region ($r<r_{+}$) must be
described by another chart. There is also a curvature singularity at
$r=r_{+}$ but this is not a problem because this is a light-like
surface thus there is no outgoing radiation. This solution, choosing
$A_{0}=0$ and $\gamma=0$, corresponds to the wormhole found in
\cite{Dotti:2006cp} with a Ricci flat base manifold.\\
 The
temperature of the four dimensional induced solution vanishes.
Finally, as expected, this solution is asymptotically locally AdS,
\textit{i.e.},
\begin{equation}
\lim_{r\rightarrow \infty} R^{\mu\nu}_{\alpha\beta} =
-\frac{1}{l^{2}}\delta^{\mu\nu}_{\alpha\beta}.
\end{equation}

The charges associated with this solution can be computed by using
Eq.(\ref{Jdiff}), in this case, for the AdS group. The background in
this case corresponds to a locally AdS space. For simplicity it is
considered the space described by the following metric
\begin{eqnarray}
g(r) =  \frac{r}{l} & &  N(r) =  C_{1}\left(\sqrt{3}-1\right)\frac{r}{l}\nonumber\\
a(r) = A_{0}, & & \Phi(r) =
C_{1}\left(\sqrt{3}-1\right)\frac{r}{l}.\label{NegativeCurvatureSolA=0}
\end{eqnarray}
The presence of $C_{1}$ is only a matter of convention. It avoids
dealing with the $\left(\sqrt{3}-1\right)$ coefficient.

As expected the electric charge, $Q(\xi)$, vanishes in this case,
since the field potential is constant. On the other hand the mass
$Q(\zeta)$, is given by
\begin{eqnarray}
M =-6\sqrt{3}\pi\kappa_{G} \frac{\gamma^{2}C_{1}^{2}}{l^{2}} V_{2},
\end{eqnarray}
where $V_{2}$ is the volume of the spatial transverse section
described by $(\theta,\phi)$.\\

\section{Conclusions and prospects}

In this work the procedure of compactification is reviewed within
the context of first order gravity. The manifest presence of four
dimensional torsion, disguised as an electric field, was well
established by Einstein himself, but in first order gravity it
becomes transparent.

Unfortunately CS gravities possess a complex phase space, therefore
many interesting solutions that one could expect to exist are not
obvious. In this article some indirect evidence of the existence of
a solution with asymptotically Reissner-Nordstr\o m behavior have
been found. This is very promising since this probably indicates
that CS gravity, a truly gauge theory, reproduces Einstein gravity
at long distances in four dimensions.

The introduction of a negative cosmological constant was also
analyzed. The results in this case are far more complex to analyze.
For simplicity the solution is constructed as perturbation over a
well known ground space \cite{Aros:2002rk}, which induces a flat
transverse section in four dimensions. The solution displayed is an
extremal black hole whose mass can be negative for a certain range
of the parameters. The existence of \textit{negative mass} solutions
is related with the presence of a negative cosmological constant,
and can not be obviously ruled out. As for $\Lambda=0$ its extension
to the non extremal case is not obvious mainly because of the non
linearity of the equations of motion.

 Torsion introduces some new degrees of freedom which has been ignored in this work, however
this is a very interesting direction to continue with this
investigation. The research for non vanishing torsion solutions into
CS gravities has proven to be a hard task. There is, however, direct
evidence \cite{Aros:2006qc} that a non vanishing torsion could give
room to reproduce standard solutions within CS gravity, in this case
standard four dimensional solutions.

%\appendix

\section*{Acknowledgments}

R.A. would like to thank Abdus Salam International Centre for
Theoretical Physics (ICTP) for its support. We also thanks
professors R. Troncoso, R. Olea and J. Oliva for valuable comments.
This work was partially funded by grants FONDECYT 1040202 and DI
06-04. (UNAB).

%\bibliographystyle{jhep}
%\bibliography{myXbib}

\providecommand{\href}[2]{#2}\begingroup\raggedright\endgroup
\end{document}